\documentclass[conference]{IEEEtran}

\usepackage{graphicx}
\usepackage{subcaption}   % side-by-side figures
\usepackage{dblfloatfix}  % improves placement of figure* in IEEEtran
\usepackage{amsmath, amssymb}
\usepackage{booktabs}
\usepackage{url}

\usepackage{tikz}
\usetikzlibrary{arrows.meta, positioning, fit, backgrounds, decorations.pathreplacing, decorations.markings, calc}

\usepackage{adjustbox}
\usepackage{xcolor}
\usepackage[section]{placeins}
\usepackage{braket}



\title{Quantum-Enhanced Processing with Tensor-Network Frontends for Privacy-Aware Federated Medical Diagnosis}

\author{
  \IEEEauthorblockN{Hiroshi Yamauchi}
  \IEEEauthorblockA{\textit{SoftBank Corp.}\\
    hiroshi.yamauchi@g.softbank.co.jp}
  \and
  \IEEEauthorblockN{Anders Peter Kragh Dalskov}
  \IEEEauthorblockA{\textit{Partisia}\\
    anderspkd@partisia.com}
  \and
  \IEEEauthorblockN{Hideaki Kawaguchi}
  \IEEEauthorblockA{\textit{Keio University}\\
    hikawaguchi@keio.jp}
  \and
    \IEEEauthorblockN{Rodney Van Meter}
    \IEEEauthorblockA{\textit{Keio University}\\
    rdv@sfc.wide.ad.jp}
}

\begin{document}
\maketitle

\begin{abstract}
We propose a privacy-aware hybrid framework for federated medical image classification that combines tensor-network representation learning, MPC-secured aggregation, and post-aggregation quantum refinement. The framework is motivated by two practical constraints in privacy-aware federated learning: MPC can introduce substantial communication overhead, and direct quantum processing of high-dimensional medical images is unrealistic with a small number of qubits. To address both constraints within a single architecture, client-side tensor-network frontends, Matrix Product State (MPS), Tree Tensor Network (TTN), and Multi-scale Entanglement Renormalization Ansatz (MERA), compress local inputs into compact latent representations, after which a Quantum-Enhanced Processor (QEP) refines the aggregated latent feature through quantum-state embedding and observable-based readout. Experiments on PneumoniaMNIST show that the effect of the QEP is frontend-dependent rather than uniform across architectures. In the present setting, the TTN+QEP combination exhibits the most balanced overall profile. The results also suggest that the QEP behaves more stably when the qubit count is sufficiently matched to the latent dimension, while noisy conditions degrade performance relative to the noiseless setting. The MPC benchmark further shows that communication cost is governed primarily by the dimension of the protected latent representation. This indicates that tensor-network compression plays a dual role: it enables small-qubit quantum processing on compressed latent features and reduces the communication overhead associated with secure aggregation. Taken together, these results support a co-design perspective in which representation compression, post-aggregation quantum refinement, and privacy-aware deployment should be optimized jointly.
\end{abstract}


\section{Introduction}

The rapid digitalization of healthcare has increased the volume of distributed, privacy-sensitive medical data stored across hospitals and institutions~\cite{rieke2020medicalfl,sheller2020brainfl}. At the same time, regulatory frameworks such as GDPR~\cite{gdpr2016}, HIPAA~\cite{hipaa1996}, APPI~\cite{appi2003}, and related medical-data governance initiatives~\cite{nextgenmedical2017,ehds2022} increasingly restrict direct data centralization and cross-institutional sharing. These constraints make collaborative learning attractive, but they also require architectures that preserve data locality and limit information exposure.

Federated learning (FL) addresses part of this requirement by allowing institutions to train joint models without transferring raw data~\cite{mcmahan2017fedavg,kairouz2021flsurvey}. However, keeping raw inputs local does not by itself eliminate privacy risk, because gradients, intermediate representations, and model updates can still leak sensitive information through inversion and inference attacks~\cite{zhu2019deepLeakage,shokri2017membership}. Secure aggregation based on multi-party computation (MPC) mitigates this problem by revealing only aggregated client contributions~\cite{bonawitz2017secureagg}. In the secret-sharing-based setting considered here, this protection is information-theoretic and therefore remains secure against computationally unrestricted adversaries, including quantum ones.

MPC strengthens privacy protection but introduces a systems-level trade-off, because protecting richer intermediate representations generally increases communication and execution cost. This trade-off becomes more restrictive when a quantum post-processing module is introduced, since the quantum branch must operate on whatever protected latent representation remains practical after secure aggregation. Meanwhile, prior work on distributed and hybrid quantum learning has studied distributed quantum models~\cite{pira2022dqnn,tang2023ceqdml,cesga2025review,ferrari2023dqc}, tensor-network/quantum hybrid architectures~\cite{chen2020hybridtn,luo2020hybridtn,qctn2024,tedq2024,quantumtrain2024,ternovaya2024tqp}, privacy-preserving quantum federated protocols~\cite{li2024ppqfl}, and quantum-secure secure-computation schemes for deep learning inference that bound leakage of both client data and model weights~\cite{sulimany2025qsmdl}. However, these lines of work have typically focused on protocol design, secure inference, or model design separately, and they do not sufficiently characterize how post-aggregation quantum refinement behaves when jointly constrained by secure aggregation and representation compression. This distinction is important for our setting: rather than replacing the federated pipeline with a fully quantum-secure inference engine, we study how representation compression and small-qubit post-aggregation quantum processing can be integrated with an MPC-based secure-aggregation model under practical deployment constraints.

A practical challenge is that current quantum processors provide only a limited number of usable qubits, making direct quantum processing of high-dimensional medical images unrealistic. A useful hybrid design must therefore compress the input before quantum processing while preserving task-relevant structure. Tensor networks (TNs) are a natural candidate for this role. Prior studies have established TNs as compact and structured learning models across supervised learning, generative modeling, and distributed settings~\cite{kardashin2020qmltn,obukhov2021spectral,chen2023residual,dborin2022mpspretrain,mossi2025mps,araz2021mpshep,zhao2024ttssm,zhang2024fltt,fhtn2024}. In the present setting, TN compression plays a dual role: it makes small-qubit quantum processing feasible on compressed latent features and reduces the communication overhead of MPC applied to those features.

Despite these promising ingredients, prior work has not sufficiently integrated representation compression, secure aggregation, and post-aggregation quantum refinement within a single privacy-aware federated pipeline. Two questions are central here: whether compressed latent representations can simultaneously improve MPC practicality and support effective small-qubit quantum processing, and whether the benefit of post-aggregation quantum refinement depends on the tensor-network frontend that produces the latent representation. 

To address these issues, we propose an end-to-end privacy-aware federated learning framework for medical image classification that combines client-side tensor-network encoding, MPC-secured aggregation of latent representations, and post-aggregation quantum refinement by a Quantum-Enhanced Processor (QEP). We compare three tensor-network frontends: Matrix Product State (MPS), Tree Tensor Network (TTN), and Multi-scale Entanglement Renormalization Ansatz (MERA), and evaluate how the effect of the QEP depends on the latent structure supplied by the frontend. The resulting design is intended to support small-qubit quantum processing while reducing the communication overhead associated with secure aggregation.

The main contributions are threefold: (1) an end-to-end privacy-aware hybrid framework combining tensor-network latent encoding, MPC-secured aggregation, and post-aggregation quantum refinement for federated medical image classification; (2) an empirical demonstration that tensor-network compression supports both small-qubit quantum processing and lower MPC communication cost; and (3) evidence that the effect of the QEP is architecture-dependent, with TTN+QEP providing the most favorable overall profile in the present setting.

The remainder of this paper is organized as follows. Section~\ref{sec:method} presents the proposed framework, including the tensor-network frontends, the MPC-secured aggregation setting, and the Quantum-Enhanced Processor. Section~\ref{sec:experimental_setup} describes the dataset, training setup, and evaluation protocol. Section~\ref{sec:experiments} defines the experimental analyses. Section~\ref{sec:results_analysis} reports the results, Section~\ref{sec:discussion} discusses their implications and limitations, and Section~\ref{sec:conclusion} concludes the paper.

\section{Methodology}
\label{sec:method}

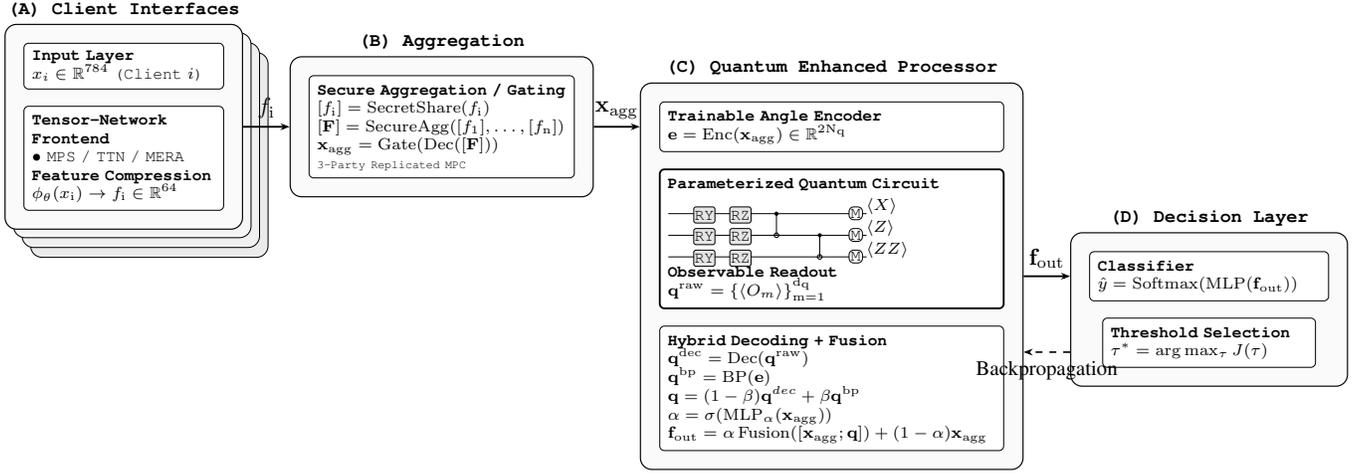
\begin{figure*}[t]
  \centering
  \begin{adjustbox}{width=\textwidth}
    % ─── スタイル定義 ───
\tikzset{
  panel/.style={draw=black, fill=black!2, rounded corners=5pt, line width=0.5pt, inner sep=8pt, text=black, font=\footnotesize\ttfamily},
  subblock/.style={draw=black, fill=white, rounded corners=2pt, line width=0.4pt, inner sep=3.5pt, text=black, font=\scriptsize\ttfamily, align=left},
  qblock/.style={draw=black, fill=white, rounded corners=2pt, line width=0.8pt, inner sep=3.5pt, text=black, font=\scriptsize\ttfamily, align=left},
  ptitle/.style={font=\footnotesize\ttfamily\bfseries, text=black},
  fwd/.style={-{Stealth[length=4pt,width=3pt]}, line width=0.8pt, color=black},
  bwd/.style={-{Stealth[length=4pt,width=3pt]}, line width=0.7pt, color=black, dashed},
  qgate/.style={draw=black, fill=black!10, minimum width=8pt, minimum height=7pt, rounded corners=1pt, font=\scriptsize\ttfamily, inner sep=0.4pt, text=black}
}

\begin{tikzpicture}[node distance=12mm]

% ─── (A) CLIENT-SIDE INTERFACES : stacked clients ─────────────
\node[subblock, text width=28mm] (a-input) {
  \textbf{Input Layer}\\
  $x_i \in \mathbb{R}^{784}$ (Client $i$)
};
\node[subblock, text width=28mm, below=2.5mm of a-input] (a-tn) {
  \textbf{Tensor-Network Frontend}\\
  $\bullet$ MPS / TTN / MERA\\
  \textbf{Feature Compression}\\
  $\phi_\theta(x_\mathrm{i})\to f_\mathrm{i} \in \mathbb{R}^{64}$
};

% back panels only (no labels inside them)
\begin{scope}[on background layer]
  \node[
    draw=black, fill=black!8, rounded corners=5pt,
    line width=0.4pt, inner sep=8pt,
    fit=(a-input)(a-tn), xshift=3.9mm, yshift=-4.3mm
  ] (panelA4) {};
  \node[
    draw=black, fill=black!6, rounded corners=5pt,
    line width=0.4pt, inner sep=8pt,
    fit=(a-input)(a-tn), xshift=2.6mm, yshift=-2.8mm
  ] (panelA3) {};
  \node[
    draw=black, fill=black!4, rounded corners=5pt,
    line width=0.4pt, inner sep=8pt,
    fit=(a-input)(a-tn), xshift=1.3mm, yshift=-1.3mm
  ] (panelA2) {};
  \node[
    panel,
    fit=(a-input)(a-tn),
    inner sep=8pt,
    label={[ptitle]north:(A) Client Interfaces}
  ] (panelA) {};
\end{scope}

%% ── (B) SECURE AGGREGATION (MPC) ─────────────────────────────
\node[subblock, text width=38mm, right=10mm of panelA.east, anchor=west] (b-mpc) {
  \textbf{Secure Aggregation / Gating}\\
$[f_\mathrm{i}] = \mathrm{SecretShare}(f_\mathrm{i})$\\
$[\mathbf{F}] = \mathrm{SecureAgg}([f_\mathrm{1}],\dots,[f_\mathrm{n}])$\\
$\mathbf{x}_\mathrm{agg} = \mathrm{Gate}(\mathrm{Dec}([\mathbf{F}]))$\\
  \textcolor{black}{\tiny 3-Party Replicated MPC}
};
\begin{scope}[on background layer]
  \node[panel, fit=(b-mpc), label={[ptitle]north:(B) Aggregation}] (panelB) {};
\end{scope}

\draw[fwd] (panelA.east) -- (panelB.west) node[midway, above, font=\normalsize] {$f_\mathrm{i}$};

%% ── (C) QUANTUM ENHANCED PROCESSOR (QEP) ──────────────────────
\node[subblock, text width=50mm, right=10mm of panelB.east, anchor=west] (c-pre) {
  \textbf{Trainable Angle Encoder}\\ 
  $\mathbf{e}=\mathrm{Enc}(\mathbf{x}_\mathrm{agg}) \in \mathbb{R}^\mathrm{2N_q}$
};

\node[qblock, text width=50mm, below=2.5mm of c-pre] (c-qengine) {
  \textbf{Parameterized Quantum Circuit}\\
  \begin{tikzpicture}[scale=0.55, baseline=(current bounding box.center)]
    \foreach \y in {0,6,12} { \draw[black, line width=0.4pt] (0,-\y mm) -- (55mm, -\y mm); }
    \node[qgate] at (10mm, 2mm) {RY}; \node[qgate] at (20mm, 2mm) {RZ};
    \node[qgate] at (10mm, -4mm) {RY}; \node[qgate] at (20mm, -4mm) {RZ};
    \node[qgate] at (10mm, -10mm) {RY}; \node[qgate] at (20mm, -10mm) {RZ};
    \draw[black, fill=black] (30mm, 0) circle (1.2pt) -- (30mm, -6mm); \draw[black] (30mm, -6mm) circle (1.8pt);
    \draw[black, fill=black] (42mm, -6mm) circle (1.2pt) -- (42mm, -12mm); \draw[black] (42mm, -12mm) circle (1.8pt);
    \foreach \y/\l in {-2/X, 4/Z, 10/ZZ} {
      \node[draw, fill=white, inner sep=0.8pt] at (52mm, -\y mm) {\scriptsize M};
      \node[right, font=\scriptsize] at (52.5mm, -\y mm) {$\langle \l \rangle$};
    }
  \end{tikzpicture}\\
  \textbf{Observable Readout}\\
$\mathbf{q}^\mathrm{raw}=\{\langle O_m \rangle\}_\mathrm{m=1}^\mathrm{d_q}$
};

\node[subblock, text width=50mm, below=2.5mm of c-qengine] (c-fusion) {
  \textbf{Hybrid Decoding + Fusion}\\
$\mathbf{q}^\mathrm{dec}=\mathrm{Dec}(\mathbf{q}^\mathrm{raw})$\\
$\mathbf{q}^\mathrm{bp}=\mathrm{BP}(\mathbf{e})$\\
$\mathbf{q}=(1-\beta)\mathbf{q}^{dec}+\beta\mathbf{q}^\mathrm{bp}$\\
$\alpha=\sigma(\mathrm{MLP}_\alpha(\mathbf{x}_\mathrm{agg}))$\\
$\mathbf{f}_\mathrm{out}=\alpha\,\mathrm{Fusion}([\mathbf{x}_\mathrm{agg};\mathbf{q}])+(1-\alpha)\mathbf{x}_\mathrm{agg}$
};

\begin{scope}[on background layer]
  \node[panel, fit=(c-pre)(c-qengine)(c-fusion), label={[ptitle]north:(C) Quantum Enhanced Processor}] (panelC) {};
\end{scope}

\draw[fwd]
  (panelB.east) -- ($(panelC.west |- panelB.east)$)
  node[midway, above, font=\normalsize] {$\mathbf{x}_\mathrm{agg}$};

%% ── (D) DECISION LAYER ───────────────────────────────────────
\node[subblock, text width=34mm, right=10mm of panelC.east, anchor=west] (d-cls) {
  \textbf{Classifier}\\
$\hat{y}=\mathrm{Softmax}(\mathrm{MLP}(\mathbf{f}_\mathrm{out}))$
};
\node[subblock, text width=30mm, below=2.5mm of d-cls] (d-loss) {
 \textbf{Threshold Selection}\\
$\mathbf{\tau}^\mathrm{*}=\arg\max_\mathrm{\tau} J(\mathbf{\tau})$\\
};
\begin{scope}[on background layer]
  \node[panel, fit=(d-cls)(d-loss), label={[ptitle]north:(D) Decision Layer}] (panelD) {};
\end{scope}

\draw[fwd]
  (panelC.east) -- ($(panelD.west |- panelC.east)$)
  node[midway, above, font=\normalsize] {$\mathbf{f}_\mathrm{out}$};

%% ── BACKWARD FEEDBACK (Adjusted Arrow) ───────────────────────
\draw[bwd]
  ($(panelD.west |- panelC.south)+(0mm,18mm)$) --
  ($(panelC.east |- panelC.south)+(0mm,18mm)$)
  node[midway, below, font=\small] {Backpropagation};

\end{tikzpicture}
  \end{adjustbox}
  \caption{Proposed TN+MPC+QEP pipeline. Client-side tensor-network encoders produce latent features, MPC-secured aggregation forms a protected global representation, and the QEP performs post-aggregation refinement before classification.}
  \label{fig:architecture}
\end{figure*}

Fig. 1 shows the proposed privacy-aware federated learning framework, which consists of client-side tensor-network encoding, MPC-secured aggregation of latent representations, and post-aggregation refinement by a Quantum-Enhanced Processor (QEP). Each client maps its local input to a compact latent representation; these latent features are securely aggregated, and the resulting global representation is refined by the QEP before final classification.

\subsection{Client-Side Tensor-Network Embedding}
\label{sec:client_tn}

Each client branch transforms an input chest X-ray image into a compact latent representation before server-side aggregation. Let
\begin{equation}
x \in \mathbb{R}^{784}
\end{equation}
denote the flattened \(28 \times 28\) grayscale image. The client-side embedding is written as
\begin{equation}
f_i = \phi_i(\psi_i(x)),
\qquad
f_i \in \mathbb{R}^{d},
\end{equation}
where \(i \in \{1,\dots,n\}\) indexes the client, \(\psi_i\) denotes a frontend-dependent preprocessing map, and \(\phi_i\) denotes the tensor-network encoder.

In the implementation, the preprocessing stage depends on the frontend type. For the MPS branch, the flattened input is first passed through a shallow real-valued block,
\begin{equation}
\psi_i^{\mathrm{MPS}} : \mathbb{R}^{784} \rightarrow \mathbb{R}^{h},
\end{equation}
consisting of a linear layer, layer normalization, and ReLU activation. The resulting feature is then partitioned into multiple sites and embedded into a complex-valued Matrix Product State (MPS) encoder with QR-based left-canonical projection.

For the TTN and MERA branches, the input is first reorganized into non-overlapping image patches. In particular, the \(28 \times 28\) image is partitioned into \(16\) patches of size \(7 \times 7\), and each patch is processed by a shared patch-wise stem network:
\begin{equation}
\psi_i^{\mathrm{tree}} : \mathbb{R}^{784}
\rightarrow
\mathbb{R}^{N_p \times d_p},
\end{equation}
where \(N_p = 16\) is the number of patches and \(d_p\) is the patch feature dimension after the stem transformation. These patch tokens are then mapped into complex local vectors and passed to either a Tree Tensor Network (TTN) or a MERA-style hierarchical encoder.

For all frontend types, the tensor-network block maps the input to a compact latent feature through a frontend-specific structured contraction. Complex internal states are used where appropriate, while QR-based projection and intermediate normalization are introduced to improve numerical stability. The final tensor-network state is converted to a real-valued vector and projected to the shared latent dimension \(d\).

\subsubsection{MPS}
For the MPS branch, the preprocessed real-valued feature
\begin{equation}
z = \psi_i^{\mathrm{MPS}}(x) \in \mathbb{R}^{h}
\end{equation}
is partitioned into a sequence of \(L\) local blocks,
\begin{equation}
z \mapsto \{z^{(1)},\dots,z^{(L)}\},
\qquad
z^{(k)} \in \mathbb{R}^{d_{\mathrm{loc}}},
\end{equation}
with \(h = L d_{\mathrm{loc}}\). Each local block is then mapped to a complex-valued local vector,
\begin{equation}
\tilde{z}^{(k)} \in \mathbb{C}^{d_{\mathrm{phys}}},
\end{equation}
through a learnable complex embedding layer. The MPS is parameterized by a sequence of site-local core tensors. At site \(k\), the core is a third-order tensor
\begin{equation}
A^{(k)} \in \mathbb{C}^{r_k \times d_{\mathrm{phys}} \times r_{k+1}},
\end{equation}
where \(d_{\mathrm{phys}}\) is the local physical dimension and \(r_k, r_{k+1}\) are the left and right bond dimensions, respectively.

Starting from an initial boundary state \(v^{(0)} \in \mathbb{C}^{r_1}\), the hidden state is updated sequentially as
\begin{equation}
v^{(k)}_{\beta}
=
\sum_{\alpha=1}^{r_k}
\sum_{s=1}^{d_{\mathrm{phys}}}
v^{(k-1)}_{\alpha}\,
A^{(k)}_{\alpha s \beta}\,
\tilde{z}^{(k)}_{s}.
\end{equation}
In the implementation, the site tensors are parameterized in a QR-projected approximately left-canonical form,
\begin{equation}
(A^{(k)})^{\dagger} A^{(k)} \approx I,
\end{equation}
and intermediate states are normalized during sequential contraction for numerical stability.

\subsubsection{TTN}
For the TTN branch, the input image is first partitioned into non-overlapping patches and mapped to a sequence of patch features,
\begin{equation}
x \mapsto \{u^{(1)},\dots,u^{(N_p)}\},
\qquad
u^{(j)} \in \mathbb{R}^{d_p},
\end{equation}
where \(N_p\) denotes the number of patches. Each patch feature is then embedded into a complex local vector
\begin{equation}
\tilde{u}^{(j)} \in \mathbb{C}^{d_{\mathrm{loc}}}.
\end{equation}
The TTN recursively combines neighboring nodes in a binary-tree fashion. Given two child states
\begin{equation}
h^{(\ell)}_{2j-1},\, h^{(\ell)}_{2j} \in \mathbb{C}^{d_{\mathrm{loc}}},
\end{equation}
their parent state is computed by concatenation followed by an isometric map,
\begin{equation}
h^{(\ell+1)}_{j}
=
\mathcal{W}^{(\ell)}
\left[
h^{(\ell)}_{2j-1}
\;\Vert\;
h^{(\ell)}_{2j}
\right],
\qquad
\mathcal{W}^{(\ell)} :
\mathbb{C}^{2d_{\mathrm{loc}}}
\rightarrow
\mathbb{C}^{d_{\mathrm{loc}}},
\end{equation}
where \([\cdot \Vert \cdot]\) denotes vector concatenation. In the implementation, \(\mathcal{W}^{(\ell)}\) is parameterized through QR-based projection so that
\begin{equation}
(\mathcal{W}^{(\ell)})^{\dagger}\mathcal{W}^{(\ell)} \approx I,
\end{equation}
and the hidden state is normalized after each aggregation level.

\subsubsection{MERA}
The MERA frontend extends the tree-structured aggregation of TTN by interleaving local mixing and coarse-graining across multiple scales. As in the TTN branch, the input image is first represented as a sequence of patch-level local features and embedded into complex local states
\begin{equation}
\tilde{u}^{(j)} \in \mathbb{C}^{d_{\mathrm{loc}}}.
\end{equation}

At each hierarchical level \(\ell\), neighboring hidden states are first transformed by disentangling operations. Given two local states
\begin{equation}
h^{(\ell)}_{a},\, h^{(\ell)}_{b} \in \mathbb{C}^{d_{\mathrm{loc}}},
\end{equation}
they are concatenated and mapped by a learned unitary transformation
\begin{equation}
\mathcal{U}^{(\ell)} :
\mathbb{C}^{2d_{\mathrm{loc}}}
\rightarrow
\mathbb{C}^{2d_{\mathrm{loc}}},
\end{equation}
followed by a split back into two updated local states. In the implementation, disentanglers are applied in alternating even and odd neighboring pairs at each level.

After local mixing, neighboring states are coarse-grained through an isometric map
\begin{equation}
\mathcal{W}^{(\ell)} :
\mathbb{C}^{2d_{\mathrm{loc}}}
\rightarrow
\mathbb{C}^{d_{\mathrm{loc}}},
\end{equation}
which produces the hidden states of the next coarser scale,
\begin{equation}
h^{(\ell+1)}_{j}
=
\mathcal{W}^{(\ell)}
\left[
\hat{h}^{(\ell)}_{2j-1}
\;\Vert\;
\hat{h}^{(\ell)}_{2j}
\right],
\end{equation}
where \(\hat{h}^{(\ell)}\) denotes the disentangled local states. Both the disentangling and coarse-graining maps are implemented through QR-projected complex transformations, and hidden states are normalized after each coarse-graining step.

\subsection{Secure Aggregation via MPC}
\label{sec:mpc}

The framework includes an MPC-secured aggregation stage between client-side tensor-network embedding and post-aggregation quantum refinement. For inference evaluation, this stage is studied as part of the end-to-end architecture in Fig.~\ref{fig:architecture}; for communication-cost analysis, it is abstracted as a protected weighted aggregation model.

Let \(f_i \in \mathbb{R}^d\) denote the latent representation produced by client \(i\), and let \(w_i \in \mathbb{R}\) denote a client-specific scalar coefficient used in the protected aggregation model. The aggregated weighted feature and total weight are defined as
\begin{equation}
\mathrm{WF} = \sum_{i=1}^{n} w_i f_i,
\qquad
\mathrm{W} = \sum_{i=1}^{n} w_i.
\end{equation}
The normalized aggregate is then written as
\begin{equation}
x = \frac{\mathrm{WF}}{\mathrm{W} + \varepsilon}.
\label{eq:normalized_agg}
\end{equation}

The MPC scenarios considered in this work correspond to progressively richer protected functionality within the abstract secure-aggregation model. Scenario~1 protects only the computation of \(\mathrm{WF}\) and \(\mathrm{W}\), and Scenario~2 additionally protects the normalization step that produces \(x\). Scenario~3 further protects the post-normalization feature-transformation stage described in Section~\ref{sec:fusion_quantum}. Scenarios~4--6 represent the active-security counterparts of Scenarios~1--3, respectively.

\subsection{Security Model and MPC Preliminaries}
\label{sec:secmodel_mpc}

\subsubsection{System Model and Threat Model}
We consider \(n\) clients \(\{C_i\}_{i=1}^n\) and an outsourced aggregation server implemented by three non-colluding computation nodes \(S_0,S_1,S_2\). Each client holds a private dataset \(\mathcal{D}_i\) and computes a local representation \(f_i \in \mathbb{R}^d\); for the MPC benchmark, a client-specific scalar coefficient \(w_i \in \mathbb{R}\) is additionally introduced to model protected weighted aggregation and normalization. We consider both passive and active security.

\subsubsection{MPC Building Blocks}
The scenarios considered later are modeled using the MPC framework of \cite{ADEN21}, together with standard techniques for secure fixed-point computation over real values \cite{DEFKSV19}.
This subsection briefly summarizes the relevant building blocks.

\paragraph{Secret sharing}
A value $v \in \mathbb{Z}_{2^k}$ is shared as follows.
Sample $v_0,v_1$ uniformly at random from \(\mathbb{Z}_{2^k}\) and define \(v_2 = v - v_0 - v_1\).
Let $[v]_i = (v_i, v_{i+1 \,\mathrm{mod}\, 3})$ denote the share held by computation node $S_i$ for $i \in \{0,1,2\}$.
In what follows, subscripts are implicitly taken modulo~3.
This sharing scheme has two key properties:
(1) any single share pair reveals no information about $v$; and
(2) any two nodes can reconstruct $v$ by combining all three components and computing \(v = v_0 + v_1 + v_2\)
This scheme is known as \emph{replicated secret sharing}.

Given secret sharings $[v]_i = (v_i, v_{i+1})$ and $[w]_i = (w_i, w_{i+1})$, the pair \((v_i + w_i,\; v_{i+1} + w_{i+1})\)
is a valid secret sharing of $v+w$.
By similar reasoning, subtraction and multiplication by public constants can also be performed locally.

To multiply \([v]\) and \([w]\), each node $S_i$ first computes
\begin{equation}
z_i = v_i w_i + v_{i+1} w_i + v_i w_{i+1}
\end{equation}
and then sends $z_i$ to $S_{i-1}$.
Each node finally defines $[z]_i = (z_i, z_{i+1})$, which is a valid secret sharing of $vw$.%
\footnote{This description omits additional correction terms required for full security. These terms do not affect the communication complexity and are therefore omitted for clarity.}
Thus, one secure multiplication requires the exchange of three $k$-bit elements.

Active security can be achieved as in \cite{ADEN21} by effectively running the protocol twice in parallel: once on the real inputs, and once on a suitably randomized encoding of those inputs.
A final consistency check is then performed to detect deviations.

\subsubsection{Fixed-Point and Truncation Semantics}
To support computation over real numbers, values are encoded in fixed-point representation.
Concretely, a real value $v \in \mathbb{R}$ is represented as
\begin{equation}
\tilde{v} = \lfloor v \cdot 2^F \rceil \in \mathbb{Z}_{2^k},
\end{equation}
where $F$ is the fractional precision parameter.
Under this encoding, multiplication yields a value scaled by $2^{2F}$, so a truncation step is required to restore the original scale.
This truncation incurs additional communication overhead: specifically, six $k$-bit elements must be exchanged \cite{DEFKSV19}.

\subsubsection{Division of Secret-Shared Values}
The normalization in \eqref{eq:normalized_agg} requires division of two secret-shared quantities, namely $\mathrm{WF}$ and $\mathrm{W}+\varepsilon$.
This can be implemented using Goldschmidt division as described in \cite{CS10}.
At a high level, the method proceeds in two stages.
First, an initial approximation to $1/(\mathrm{W}+\varepsilon)$ is computed.
Second, an iterative refinement is performed to approximate $\mathrm{WF}/(\mathrm{W}+\varepsilon)$.
The first stage requires a truncation step and a secure computation of the effective bit width of $\mathrm{W}+\varepsilon$.
Determining the bit width of a secret-shared value requires bit decomposition, which can be implemented using approximately $k$ secure multiplications.
The second stage requires two fixed-point multiplications per iteration.
The required number of iterations depends on the ring size.

\subsubsection{Cost Model}
We model secure aggregation using replicated secret sharing over \(\mathbb{Z}_{2^k}\) with three non-colluding computation nodes. Under this model, addition and multiplication by public constants are local, whereas secure multiplication requires communication between nodes. Real-valued computation is handled in fixed-point form, so multiplication additionally requires truncation, and normalization requires secure division.

In the communication model used in this paper, one secure multiplication costs \(3k\) bits, one truncation costs \(6k\) bits, and a full fixed-point multiplication therefore costs \(9k\) bits. Secure division is modeled with communication cost \(3k(k+4\theta+2)\), where \(\theta\) denotes the number of refinement iterations. Active security is modeled by doubling the communication cost. Because the focus here is communication benchmarking rather than protocol derivation, we use these established costs directly.

\subsection{Quantum-Enhanced Processor}
\label{sec:fusion_quantum}

The Quantum-Enhanced Processor (QEP) refines the aggregated latent representation through quantum-state embedding, observable-based readout, and classical fusion. Let the aggregated server-side feature be denoted by
\begin{equation}
\mathbf{x}_{\mathrm{agg}} \in \mathbb{R}^{d}.
\end{equation}
In the present implementation, the quantum circuit is evaluated by statevector simulation and used as a fixed nonlinear feature transformation rather than as a trainable variational component. Accordingly, only the surrounding classical modules, including the encoder, decoder, bypass pathway, and fusion layers, are optimized during training.

From a representational viewpoint, the QEP can be interpreted as a feature-expansion mechanism that embeds a compact latent vector into a quantum Hilbert space and extracts observable-based statistics. In this setting, the effective dimensionality of the quantum feature space is determined by the number of observables used for measurement rather than by the number of circuit parameters. To ensure sufficient expressivity, it is therefore desirable that the number of quantum features \(d_q\) be comparable to the latent dimension \(d\). When using both one-body and two-body Pauli observables, this leads to a scaling relation \(d_q = \mathcal{O}(N_q^2)\), suggesting that the number of qubits should scale approximately as \(N_q \sim \sqrt{d}\). This consideration motivates the qubit counts used in the present study.

\subsubsection{Quantum Encoding and Observable Readout}

\begin{figure}[!t]
    \centering
    \includegraphics[width=0.48\textwidth]{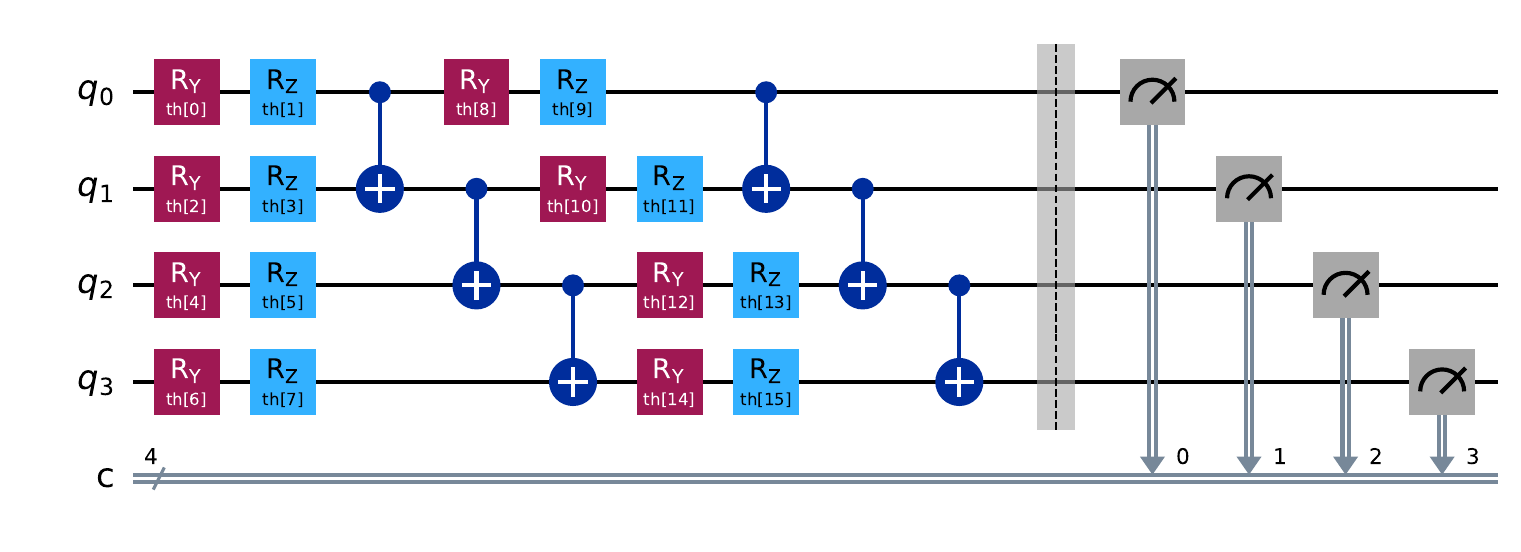}
    \caption{
    Parameterized quantum circuit used in the QEP.
    Each layer applies single-qubit \(R_y\) and \(R_z\) rotations followed by nearest-neighbor CNOT entangling gates.
    For visualization, the circuit diagram is shown in a 4-qubit instance, and terminal measurement gates are drawn explicitly.
    In the actual implementation, however, the circuit output is summarized through expectation values of selected Pauli observables rather than bitstring sampling.
    In the experiments, the same circuit template was evaluated with different qubit counts, up to \(N_q=16\).
    }
    \label{fig:pqe_circuit}
\end{figure}

The QEP first applies a trainable angle encoder
\begin{equation}
\mathbf{e} = \mathrm{Enc}(\mathbf{x}_{\mathrm{agg}}),
\qquad
\mathbf{e} \in \mathbb{R}^{2N_q},
\end{equation}
where \(N_q\) is the number of qubits and \(\mathrm{Enc}(\cdot)\) denotes a shallow multilayer perceptron with normalization and nonlinear activation. For each layer \(l\) and qubit \(q\), the rotation angles are defined as
\begin{equation}
\theta^{(l,q)}_{y}
=
\pi s \left(e_{2q-1} + \delta^{(l,q)}_{y}\right),
\qquad
\theta^{(l,q)}_{z}
=
\pi s \left(e_{2q} + \delta^{(l,q)}_{z}\right),
\end{equation}
where \(s\) is a global scaling factor and \(\delta^{(l,q)}_{y}, \delta^{(l,q)}_{z}\) are circuit-side offsets.

Starting from \(\ket{0}^{\otimes N_q}\), the circuit applies \(L\) repeated layers of single-qubit rotations followed by nearest-neighbor entangling gates:
\begin{equation}
\begin{aligned}
U(\mathbf{x}_{\mathrm{agg}})
&=
\prod_{l=1}^{L}
\Biggl[
\Bigl(\prod_{q=1}^{N_q-1}\mathrm{CX}_{q,q+1}\Bigr)
\\
&\qquad\qquad
\Bigl(\prod_{q=1}^{N_q} R_z\!\left(\theta^{(l,q)}_{z}\right)
R_y\!\left(\theta^{(l,q)}_{y}\right)\Bigr)
\Biggr].
\end{aligned}
\end{equation}

For a circuit with \(N_q\) qubits and depth \(L\), this construction requires \(2LN_q\) single-qubit rotation gates and \(L(N_q-1)\) nearest-neighbor CNOT gates. Fig.~\ref{fig:pqe_circuit} illustrates the circuit structure used in the QEP.

This leads to an overall gate complexity of \(\mathcal{O}(L N_q)\), with nearest-neighbor entangling gates. In contrast, when both one-body and two-body observables are included, the effective quantum feature dimension scales as \(d_q = \mathcal{O}(N_q^2)\). Thus, the circuit realizes a quadratic feature expansion while maintaining linear gate complexity.

The output quantum state is summarized through expectation values of Pauli observables. For each observable \(\hat{O}_m\), the corresponding quantum statistic is defined as
\begin{equation}
q_m^{\mathrm{raw}}
=
\bra{\psi(\mathbf{x}_{\mathrm{agg}})}
\hat{O}_m
\ket{\psi(\mathbf{x}_{\mathrm{agg}})},
\end{equation}
yielding a feature vector
\begin{equation}
\mathbf{q}^{\mathrm{raw}} \in \mathbb{R}^{d_q},
\end{equation}
where \(d_q\) is determined by the chosen observable set. This observable-based readout enables extraction of nonlinear quantum features from a compact latent representation, as only expectation values of selected observables are evaluated. This construction should be interpreted as an observable-based quantum feature map rather than a variational, kernel-based, or tomography-based approach.

\subsubsection{Classical Decoding and Residual Fusion}
The raw quantum statistics are first decoded into the latent feature space through a trainable decoder
\begin{equation}
\mathbf{q}^{\mathrm{dec}}
=
\mathrm{Dec}\!\left(\mathbf{q}^{\mathrm{raw}}\right)
\in \mathbb{R}^{d}.
\end{equation}
In parallel, the encoded angle vector is passed through a differentiable bypass pathway,
\begin{equation}
\mathbf{q}^{\mathrm{bp}}
=
\mathrm{BP}(\mathbf{e})
\in \mathbb{R}^{d},
\end{equation}
and the two branches are mixed as
\begin{equation}
\mathbf{q}
=
(1-\beta)\,\mathbf{q}^{\mathrm{dec}}
+
\beta\,\mathbf{q}^{\mathrm{bp}},
\end{equation}
where \(\beta \in (0,1)\) is a learned scalar gate. This design allows the encoder to remain trainable even though the quantum simulation itself is treated as non-differentiable.

The resulting quantum-enhanced representation is then fused with the original classical feature:
\begin{equation}
\mathbf{z}
=
\mathrm{Fusion}\!\left([\mathbf{x}_{\mathrm{agg}};\mathbf{q}]\right).
\end{equation}
A sample-wise scalar gate
\begin{equation}
\alpha
=
\sigma\!\left(
W_{\alpha}\,\mathrm{LN}(\mathbf{x}_{\mathrm{agg}})+b_{\alpha}
\right)
\end{equation}
controls the final interpolation
\begin{equation}
\mathbf{f}_{\mathrm{out}}
=
\alpha\,\mathbf{z}
+
(1-\alpha)\,\mathbf{x}_{\mathrm{agg}}.
\end{equation}

\subsubsection{Training Objective}
The final classifier is trained with weighted cross-entropy. In addition, an auxiliary supervision term can be applied to the representation produced by the quantum-enhanced branch:
\begin{equation}
\mathcal{L}
=
\mathcal{L}_{\mathrm{cls}}
+
\lambda_q \mathcal{L}_{\mathrm{aux}},
\end{equation}
where \(\mathcal{L}_{\mathrm{aux}}\) encourages the quantum-enhanced pathway to retain task-relevant information. Overall, the QEP should be interpreted as a hybrid feature-refinement module whose contribution depends on how quantum-derived features interact with the latent structure provided by the frontend architecture.

\section{Experimental Setup}
\label{sec:experimental_setup}

We evaluated the proposed framework on \texttt{PneumoniaMNIST}, a binary medical image classification benchmark from the MedMNIST collection~\cite{medmnist}. The task is to discriminate \textit{Normal} from \textit{Pneumonia} using resized \(28 \times 28\) grayscale chest X-ray images normalized to \([0,1]\). For the MPS branch, each image was flattened into a 784-dimensional vector. For the TTN and MERA branches, the same image was additionally reorganized into \(16\) non-overlapping patches of size \(7 \times 7\), followed by a shared patch-wise stem transformation. The standard training, validation, and test splits provided by the dataset interface were used throughout.

\begin{figure}[t]
    \centering
    \includegraphics[width=0.95\linewidth]{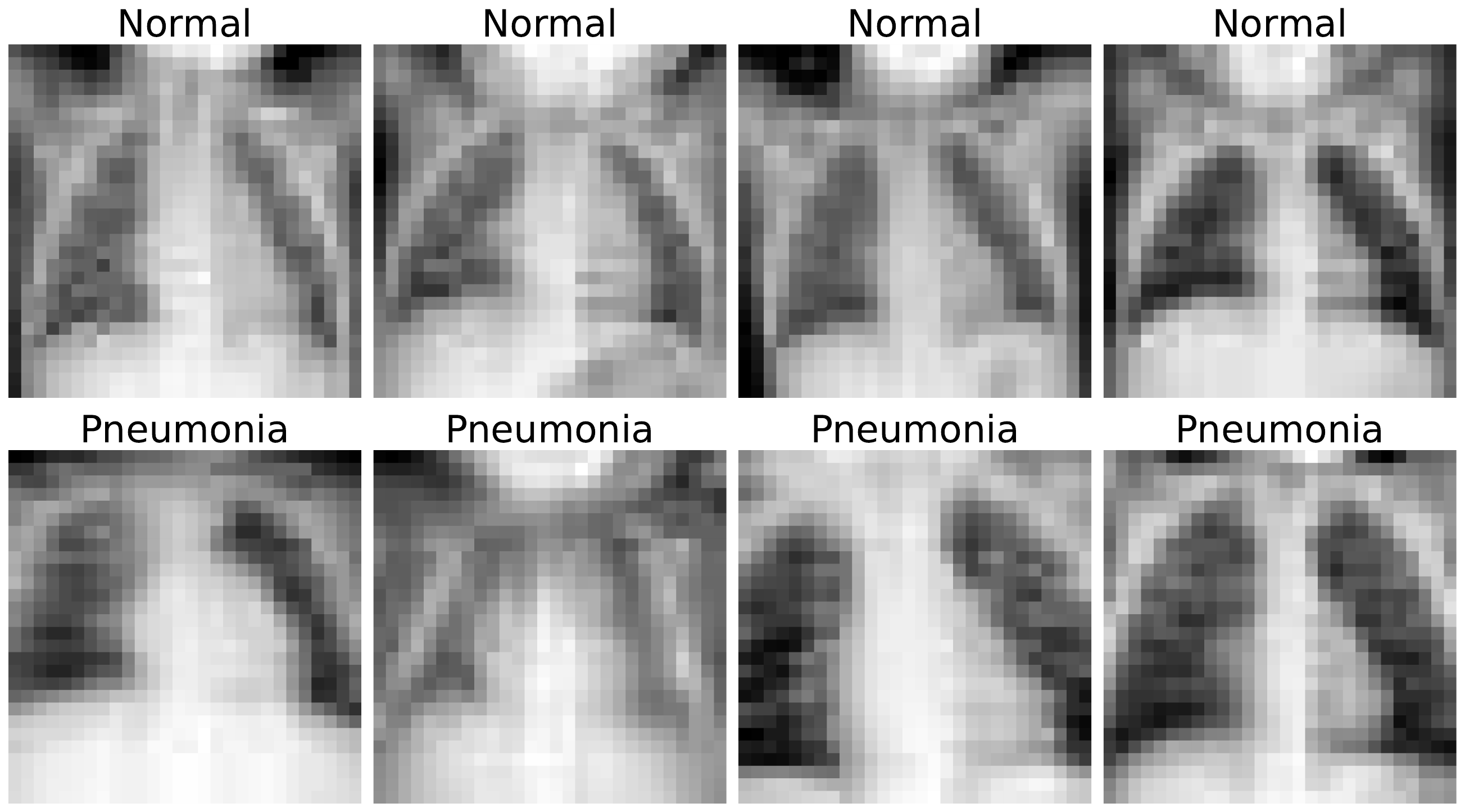}
    \caption{Example samples from PneumoniaMNIST. The top row shows Normal cases and the bottom row shows Pneumonia cases.}
    \label{fig:pneumoniamnist_samples}
\end{figure}

To reflect the distributed setting, each split was partitioned across client branches using label-stratified client assignment. Unless otherwise noted, the number of client branches was fixed at 16. Training used a per-client local batch size of 4, yielding up to 64 samples per federated training step when all client branches contributed non-empty batches.

The overall model consisted of client-side tensor-network embedding, server-side feature gating, and an optional Quantum-Enhanced Processor (QEP), followed by a classical classifier. We considered three tensor-network frontends: Matrix Product State (MPS), Tree Tensor Network (TTN), and Multi-scale Entanglement Renormalization Ansatz (MERA). In all cases, the shared latent feature dimension was fixed to 64.
The comparison is intended to evaluate practically instantiated frontend families under a common latent dimension, optimization protocol, and evaluation setting, rather than a strictly topology-isolated control.

Training was performed for 20 epochs using Adam with parameter-group-specific learning rates and weight decay \(10^{-5}\). The tensor-network, patch-stem, quantum-encoder-related, and remaining head parameters were optimized with learning rates \(1\times10^{-5}\), \(3\times10^{-5}\), \(5\times10^{-5}\), and \(1\times10^{-4}\), respectively. To account for class imbalance, weighted cross-entropy was used with class weights computed from the training split. In the quantum-enabled setting, the auxiliary supervision term for the quantum-enhanced branch was activated with weight 0.5.

For the QEP, the quantum branch used 16 qubits, circuit depth 2, angle scale 0.5, and refresh interval 1. Quantum expectation values were computed with Qiskit Aer in statevector mode. The circuit used repeated layers of single-qubit \(R_y\) and \(R_z\) rotations followed by nearest-neighbor CNOT entangling gates, and the observable set included one-body Pauli terms together with selected two-body correlations. The resulting quantum statistics were classically decoded into the shared latent feature space and adaptively fused with the classical pathway.

We report both standard-threshold and threshold-optimized evaluation. Performance was assessed using Accuracy, Precision, Recall, and F1-score, and quantum runs additionally tracked internal QEP diagnostics including the mean fusion coefficient $\alpha$ and the standard deviation of the quantum-enhanced branch representation.

We also examined QEP behavior under qubit scaling and noise variation. Qubit counts were varied from 4 to 16, and noise sensitivity was evaluated at 8 qubits under noiseless, depolarizing, thermal, and mixed-noise conditions. These analyses should be interpreted as controlled studies of operating range and robustness rather than hardware-performance claims.

Based on this common setting, we define two complementary experiments: one for end-to-end predictive analysis and one for MPC communication-cost benchmarking.
Within the end-to-end analysis, additional qubit-scaling and noise studies focus on the TTN-based model because the frontend comparison identifies TTN+QEP as the most balanced configuration in the present setting. This choice is intended to characterize the operating range of the quantum branch without implying that the same range transfers unchanged to other frontend types.

% ------------------------------------------------------------------------------
%  Section: Experiments
% ------------------------------------------------------------------------------
\section{Experiments}
\label{sec:experiments}

\subsection{Experiment 1: Quantum Enhancement across MPS, TTN, and MERA}
\label{sec:exp1}

The first experiment examines whether the effect of the Quantum-Enhanced Processor (QEP) depends on the client-side tensor-network frontend. We compare Classical and Quantum modes for MPS, TTN, and MERA under the common training and evaluation setting described in Section~\ref{sec:experimental_setup}.

In the Classical mode, the server-side latent feature is passed directly to the classifier. In the Quantum mode, the same latent feature is additionally processed by the QEP, which produces observable-based quantum statistics, decodes them into the shared latent space, and fuses them with the classical representation. This comparison isolates the contribution of post-aggregation quantum refinement under matched conditions.

We report both predictive performance and internal QEP behavior. Predictive evaluation uses both the standard threshold and the validation-optimized threshold. Internal analysis tracks the mean fusion coefficient \(\alpha\), the dispersion of the quantum-enhanced branch through \(q\)-std, and their training behavior across frontend types.

Because TTN provides the most favorable overall profile in the frontend comparison, we additionally use the TTN-based model to examine the operating range of the QEP under qubit scaling and noise variation. For qubit scaling, the QEP is instantiated with \(N_q \in \{4,6,8,10,12,14,16\}\), and the resulting test-accuracy distributions are compared with the classical TTN baseline. For noise sensitivity, we fix \(N_q = 8\) and compare the classical TTN baseline with four quantum execution conditions: noiseless, depolarizing, thermal, and mixed noise. Both analyses are reported as distributional summaries of test accuracy over repeated runs.

\subsection{Experiment 2: MPC Mode Benchmark}
\label{sec:exp2}

The second experiment benchmarks the communication cost of MPC-secured aggregation under progressively stronger protection settings. The benchmark isolates the communication induced by protected latent aggregation and is not intended as a full accounting of all end-to-end distributed traffic. Rather, it is designed to clarify how protected representation dimension and security setting determine the communication burden of the secure aggregation stage.

We vary the number of clients from \(n=1\) to \(30\). For each value of \(n\), we consider seven scenarios. Scenario~0 is an insecure baseline in which a central server \(S\) performs the aggregation and transformation directly. Scenarios~1--3 correspond to passive-security MPC settings with progressively richer protected functionality: Scenario~1 protects only weighted aggregation, Scenario~2 additionally protects normalization, and Scenario~3 further protects the post-aggregation transformation stage. Scenarios~4--6 are the corresponding active-security counterparts. Table~\ref{tab:mpc_scenarios} summarizes these settings.

\begin{table}[t]
\caption{Scenario definitions for the MPC benchmark.}
\label{tab:mpc_scenarios}
\centering
\footnotesize
\begin{tabular}{cl}
\toprule
Scenario & Definition \\
\midrule
0 & Insecure baseline \\
1 & Passive, aggregation only \\
2 & Passive, aggregation + normalization \\
3 & Passive, aggregation + normalization + transformation \\
4 & Active, aggregation only \\
5 & Active, aggregation + normalization \\
6 & Active, aggregation + normalization + transformation \\
\bottomrule
\end{tabular}
\end{table}

\subsubsection{Cost Model and Measurement Protocol}

We employ a symbolic cost meter that tracks three types of communication: (i) data sent by clients to computation nodes, (ii) data exchanged between computation nodes, and (iii) data transmitted during output reconstruction.

In all scenarios, we assume \(k=64\), corresponding to an encoding in which client inputs fit into 64-bit machine words. For Scenarios~2 and~5, division is metered using \(\theta = 5\). The benchmark is intended to capture relative scaling trends across MPC scenarios rather than protocol-tight cryptographic cost bounds.

\subsubsection{Fairness and Comparability Across Modes}

To avoid conflating security modeling with numerical instability, we assume that fixed-point parameters can be chosen so that the accuracy of the MPC computation matches that of the original insecure floating-point computation. As noted in \cite{RRGGSCR21}, automated tools can assist in fixed-point parameter selection. Under this assumption, the benchmark should be interpreted as a comparison of secure-computation overheads under matched numerical behavior.

\section{Results and Analysis}
\label{sec:results_analysis}

\subsection{Results of Experiment 1: Quantum Enhancement across MPS, TTN, and MERA}
\label{sec:results_exp1}

\begin{figure}[t]
    \centering
    \includegraphics[width=0.49\textwidth]{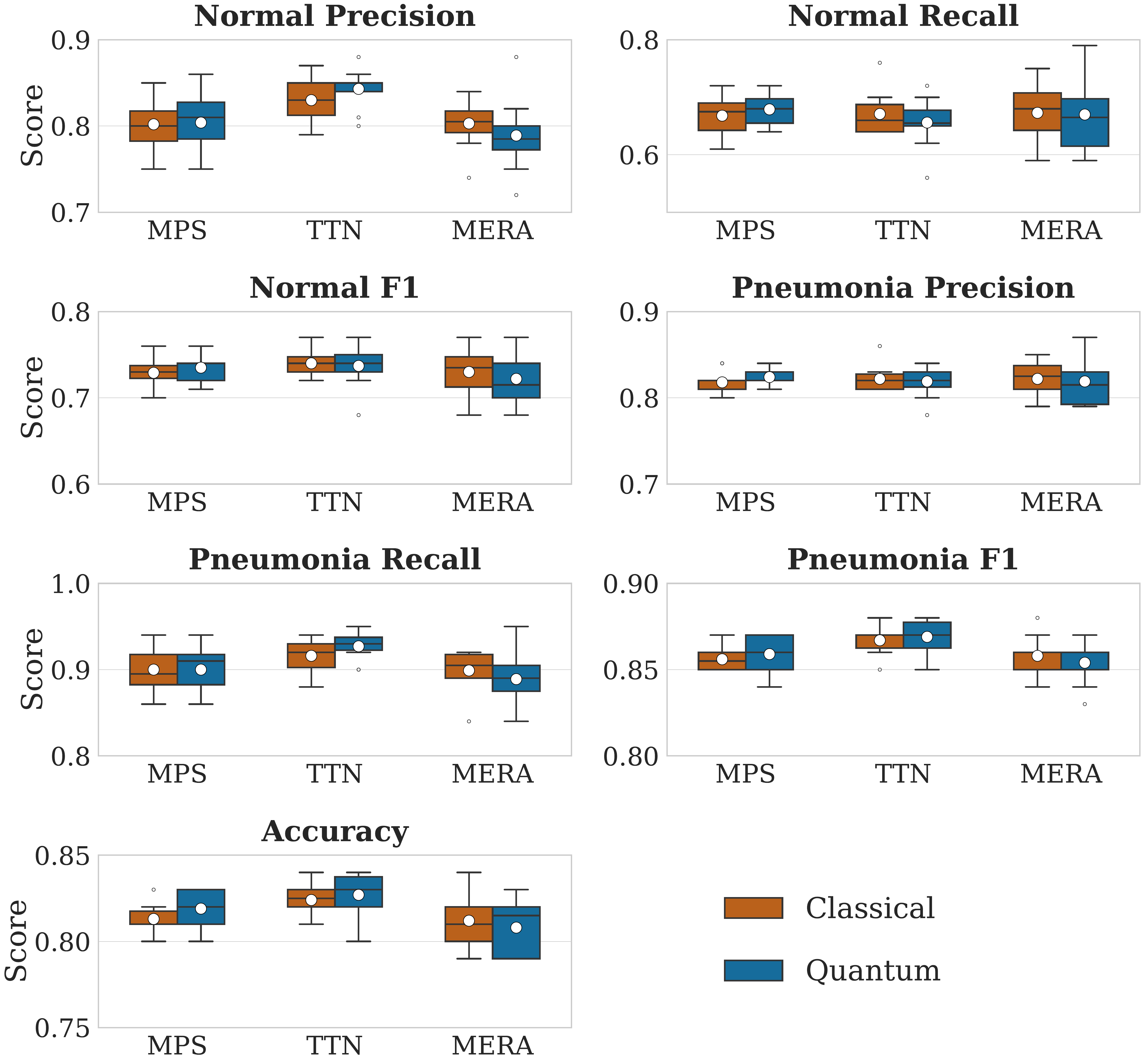}
    \caption{Threshold-optimized test performance of the Classical and Quantum modes across MPS, TTN, and MERA. Boxplots report class-wise Precision, Recall, F1-score, and overall Accuracy for the default QEP setting (\(N_q=16\)).}
    \label{fig:optimized_metrics}
\end{figure}

\begin{figure}[t]
    \centering
    \includegraphics[width=0.46\textwidth]{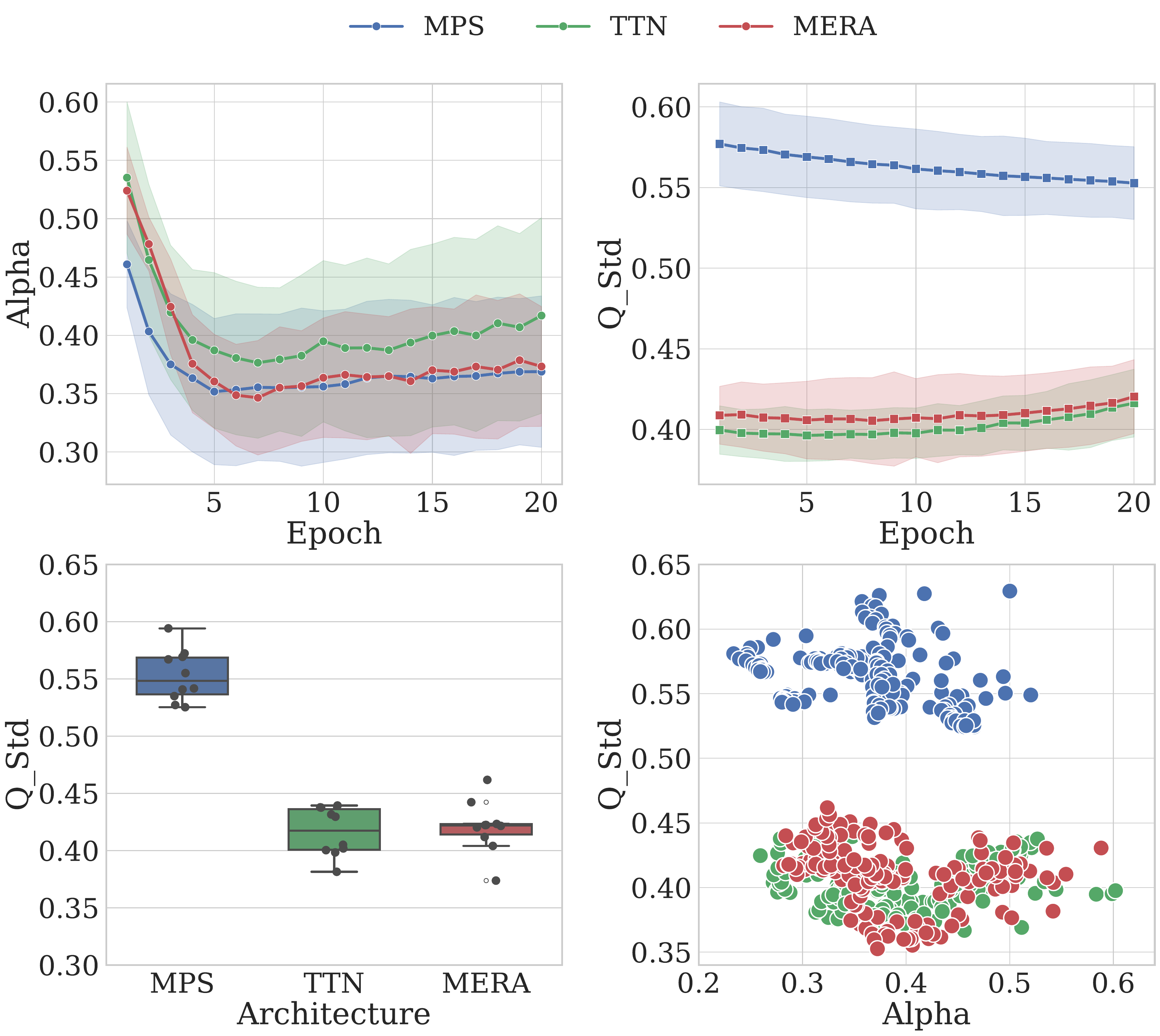}
    \caption{Internal behavior of the QEP across MPS, TTN, and MERA: (A) evolution of \(\alpha\), (B) evolution of \(q\)-std, (C) final-epoch \(q\)-std distribution, and (D) \(\alpha\)-\(q\)-std phase behavior.}
    \label{fig:qep_dynamics}
\end{figure}

\begin{figure}[t]
    \centering
    \begin{subfigure}[b]{0.68\linewidth}
        \centering
        \includegraphics[width=\linewidth]{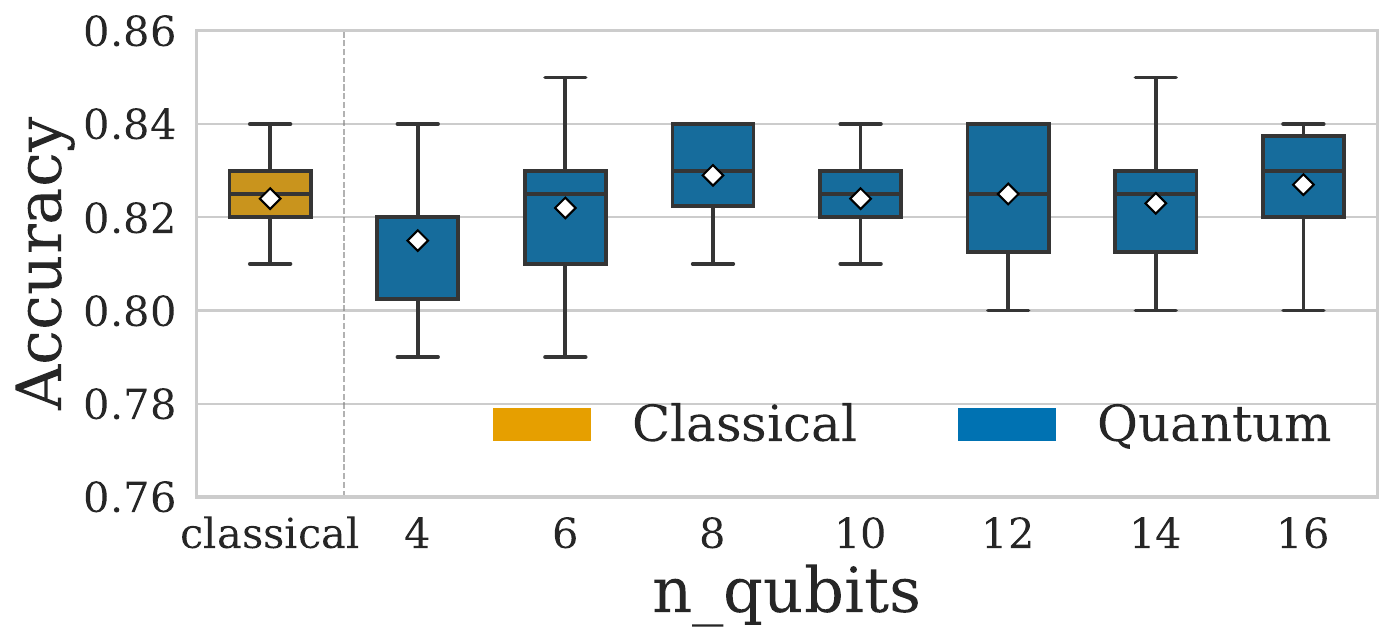}
        \caption{Qubit-count scaling (\(N_q=4\) to \(16\)).}
        \label{fig:qep_nqubit_scaling}
    \end{subfigure}
    \hfill
    \begin{subfigure}[b]{0.66\linewidth}
        \centering
        \includegraphics[width=\linewidth]{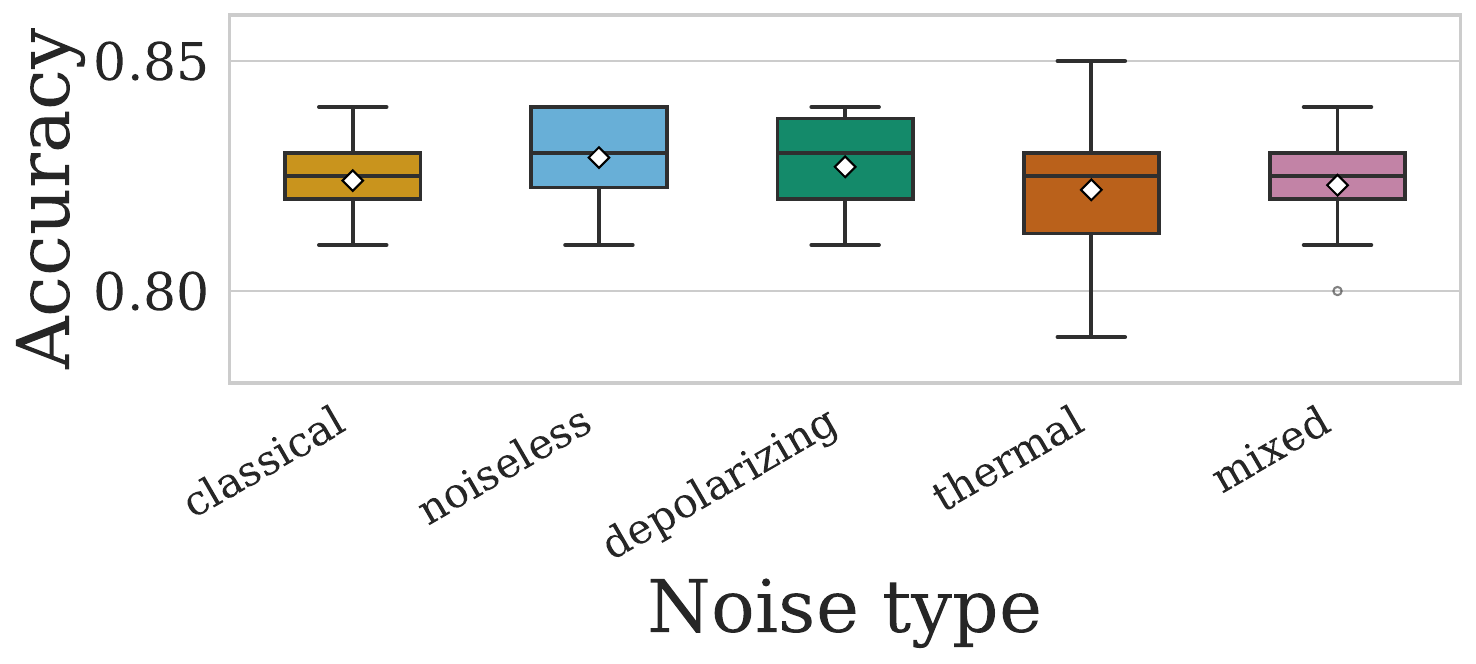}
        \caption{Noise robustness at \(N_q=8\).}
        \label{fig:qep_noise_n8}
    \end{subfigure}
    \caption{TTN-based QEP: (a) test-accuracy distributions under qubit-count scaling and (b) robustness under representative noise conditions at \(N_q=8\). The classical TTN baseline is shown for reference.}
    \label{fig:qep_scalability_noise}
\end{figure}

Figs.~\ref{fig:optimized_metrics} and~\ref{fig:qep_dynamics} summarize the comparison between Classical and Quantum modes across MPS, TTN, and MERA. Overall, the results show that the effect of quantum enhancement is architecture-dependent rather than uniform across tensor-network frontends.

Under threshold-optimized evaluation, TTN+QEP provides the most balanced overall profile among the tested frontend configurations. Across architectures, the main effect of the QEP is not a uniform increase in Accuracy, but a frontend-dependent redistribution of class-wise Precision, Recall, and F1. Pneumonia Recall remains high across all three frontend types, staying around 0.90 in the present setting.

The internal diagnostics also differ across frontends. The mean fusion coefficient \(\alpha\) decreases during training for all architectures, but the resulting operating regimes are distinct. MPS reaches the highest final \(q\)-std, MERA remains intermediate, and TTN stabilizes at a comparatively lower and more controlled quantum-branch dispersion.
These observations suggest that the effect of the QEP depends not only on the quantum branch itself, but also on the latent structure induced by the tensor-network frontend.

Fig.~\ref{fig:qep_scalability_noise} shows additional analyses for the TTN-based QEP. In the qubit-scaling comparison, the lower-qubit settings, especially \(N_q=4\) and \(N_q=6\), tend to show less favorable accuracy distributions than the higher-qubit settings. By contrast, from \(N_q \ge 8\), the distributions appear more stable and remain within a broadly similar range.

This behavior is qualitatively consistent with the scaling consideration introduced in Section~\ref{sec:fusion_quantum}. For the present latent dimension \(d=64\), the transition to more stable behavior from around \(N_q=8\) is compatible with the heuristic relation \(N_q \sim \sqrt{d}\). Although this does not constitute a formal scaling law, it supports the view that the usefulness of the QEP depends on a reasonable match between quantum feature capacity and latent representation size.

The noise analysis at \(N_q=8\) shows degradation under all tested noisy conditions relative to the noiseless setting. The depolarizing condition remains comparatively close to the noiseless case, whereas the thermal condition shows a wider spread and includes lower-performing runs. The mixed-noise condition also shows a degradation trend while retaining a broadly comparable range.

\subsection{Results of Experiment 2: MPC Benchmark}
\label{sec:exp2_results}

\begin{figure}[t]
    \centering
    \begin{subfigure}[b]{\linewidth}
        \centering
        \includegraphics[width=\linewidth]{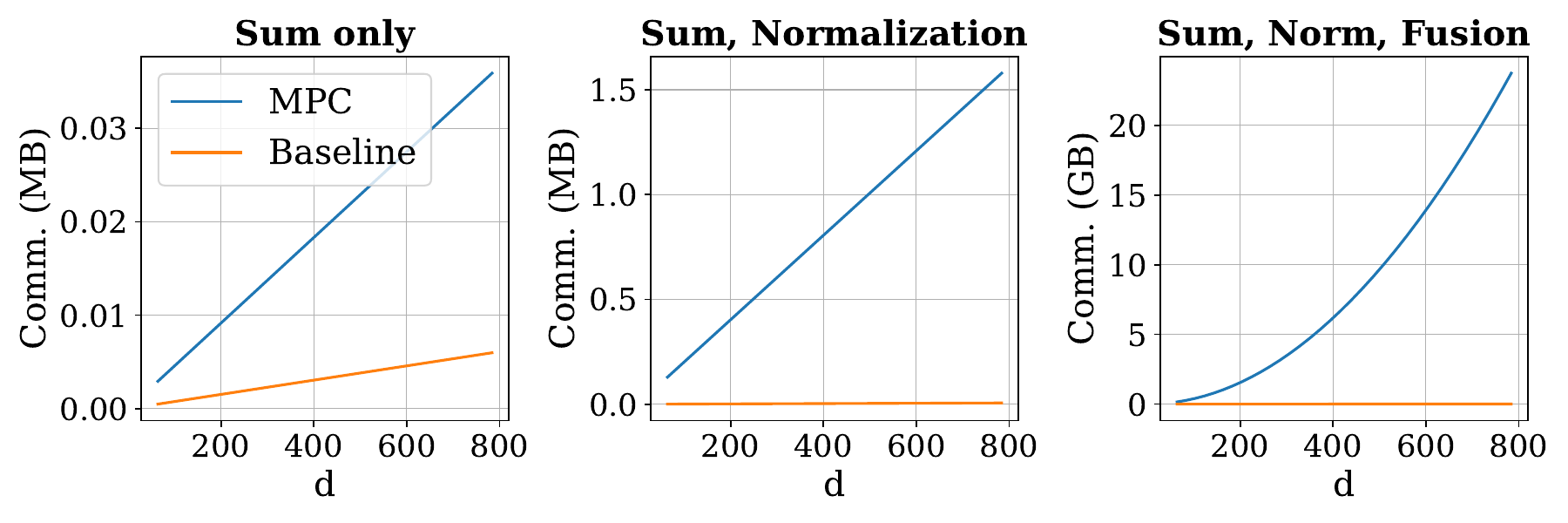}
        \caption{Passive MPC scenarios (Scenarios 1--3).}
        \label{fig:comm_passive}
    \end{subfigure}
    
    \vspace{0.5em}
    
    \begin{subfigure}[b]{\linewidth}
        \centering
        \includegraphics[width=\linewidth]{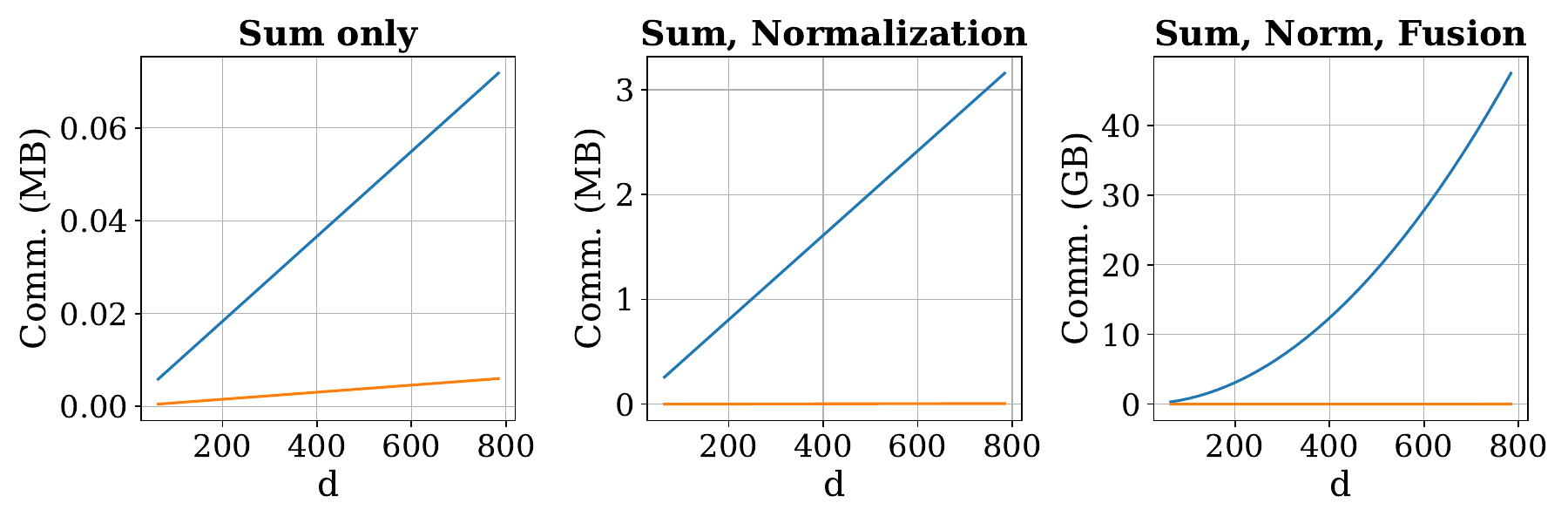}
        \caption{Active MPC scenarios (Scenarios 4--6).}
        \label{fig:comm_active}
    \end{subfigure}
    \caption{Modeled communication overhead as a function of protected representation dimension for passive and active MPC scenarios.}
    \label{fig:comm_mpc}
\end{figure}

Fig.~\ref{fig:comm_passive} and~\ref{fig:comm_active} summarize the modeled communication overhead for the passive and active MPC scenarios together with the insecure baseline.

The benchmark shows that, under the present abstraction of server-side MPC communication, the dominant communication cost is governed by the bit-width \(k\), the protected representation dimension, and the selected protection scenario, rather than by the number of aggregated clients itself. In both figures, \(k=64\) is fixed while the protected representation dimension varies from the raw dimension \(D_{\mathrm{raw}}=784\) to the compressed latent dimension \(d=64\).

Active security preserves the same qualitative scaling trend as passive security while introducing an additional multiplicative overhead. Across both passive and active settings, the dominant communication trend is driven by the representation dimension, indicating that client-side compression directly reduces the communication surface on which secure aggregation is applied.

Because the present benchmark isolates the communication cost of the MPC stage itself, the number of clients does not emerge as the dominant scaling factor in the reported curves. Under the linear secret-sharing model considered here, the main cost trend is governed primarily by protected representation dimension and protection scenario, indicating that client-side compression has direct systems-level value because it reduces the communication surface on which secure aggregation is applied.

\section{Discussion}
\label{sec:discussion}

The results of Experiment~1 suggest that the QEP should be understood not as a universal accuracy booster, but as a quantum feature-refinement mechanism whose effectiveness depends on how it is paired with the tensor-network frontend. Across MPS, TTN, and MERA, the QEP does not yield a uniform increase in Accuracy or F1-score; instead, its contribution appears through architecture-dependent changes in class-wise operating characteristics and internal fusion dynamics. The main implication is therefore not the superiority of any single component in isolation, but the importance of jointly designing tensor-network structure and post-aggregation quantum refinement.

Within this broader picture, TTN exhibits the most favorable overall profile, achieving strong predictive performance, including high Accuracy and Pneumonia F1, while maintaining high sensitivity to pneumonia-positive cases. More importantly, the results indicate that the contribution of the QEP is conditioned by the latent representation supplied by the frontend. MPS, TTN, and MERA do not merely compress the input differently; they expose different latent organizations to the QEP, leading to distinct quantum operating regimes.

The internal diagnostics support this interpretation. MPS exhibits the largest quantum-branch dispersion, whereas TTN remains in a lower and more controlled \(q\)-std regime, suggesting that larger internal quantum variability does not automatically translate into stronger downstream performance. In the present experiment, the most favorable behavior emerges not from the largest quantum-branch spread, but from a more balanced interaction between quantum refinement and the frontend-induced latent structure. From a quantum-machine-learning perspective, this indicates that the effect of post-aggregation quantum enhancement is jointly determined by the quantum map and the representational topology that feeds it.

The qubit-scaling and noise analyses provide an additional practical perspective on the TTN-based QEP. For the present 64-dimensional latent input, lower-qubit configurations, particularly \(N_q=4\) and \(N_q=6\), appear to provide insufficient feature capacity, whereas behavior becomes more stable from \(N_q=8\) onward. In addition, all noisy conditions at \(N_q=8\) show some degradation relative to the noiseless setting, although the extent depends on the noise model. These observations do not establish hardware scalability, but indicate that the hybrid pipeline retains meaningful behavior once the quantum register size is reasonably matched to the latent dimension.

From a clinical perspective, the consistently high Pneumonia Recall across architectures is encouraging. In screening or triage settings, sensitivity to pneumonia-positive cases is often more important than small fluctuations in class-balanced metrics because false negatives carry disproportionate clinical cost.

At the same time, the present results should not be read as establishing TTN+QEP as a universally optimal combination. Rather, they indicate that the utility of post-aggregation quantum refinement depends on the latent structure delivered by the frontend and its alignment with the effective capacity of the quantum feature map. In this sense, the main lesson is not the dominance of a single architecture, but the importance of co-design across compression, aggregation, and quantum refinement.

Experiment~2 complements this interpretation from the systems side. Because communication cost is dominated by the protected representation dimension, tensor-network compression directly reduces the communication overhead incurred by MPC on that representation, improving the practicality of privacy-preserving deployment.

A related direction for quantum feature extraction in this framework is to replace or augment the present observable-based QEP with a quantum reservoir readout acting on the aggregated latent input. Quantum reservoir computing is attractive in this context because it exploits the natural nonlinear dynamics of many-body quantum systems while avoiding gradient-based variational optimization of the quantum circuit itself. This is relevant from a practical viewpoint, as variational quantum models can suffer from barren plateau phenomena, whereas quantum kernel methods can also face exponential concentration of kernel values and associated measurement inefficiency under certain conditions~\cite{larocca2025barren,thanasilp2024kernel}. By contrast, reservoir-style quantum processing uses the dynamics of a fixed quantum system as a feature generator and trains only the classical readout, providing a complementary design point for small- to intermediate-scale hybrid learning. Recent studies have experimentally demonstrated repeated-measurement quantum reservoir computing on superconducting devices and large-scale analog quantum reservoir learning at the 100-qubit scale, supporting the practical relevance of this direction~\cite{yasuda2023qrc_repeated}.

The longer-term significance of the architecture lies in future qubit regimes where faithful classical simulation is no longer practical, in which case post-aggregation quantum processing on tensor-network-compressed latent features becomes more meaningful than direct processing of the original high-dimensional input. This interpretation is consistent with the qubit-scaling result, suggesting that frontend structure and quantum resource scale should be considered jointly.

At the same time, the present study has several limitations. First, the quantum branch is evaluated through Qiskit Aer statevector simulation, so the results should be interpreted as evidence for hybrid representational behavior rather than as a demonstration of hardware-level feasibility. Second, the present study does not establish that the quantities computed by the QEP are classically intractable or impractical to obtain classically; rather, the QEP is used as a structured small-qubit feature map. Third, the current comparison does not yet isolate which portion of the observed QEP effect is uniquely attributable to the quantum observable map itself, and the MPC benchmark remains a symbolic communication-cost model rather than a full protocol-level implementation.

\section{Conclusion}
\label{sec:conclusion}

In this work, we investigated the role of the QEP within a privacy-aware federated learning framework for medical image classification. The proposed architecture combines client-side tensor-network representation learning, an abstract secure-aggregation model for MPC benchmarking, and a post-aggregation quantum refinement module.

Our results show that the contribution of the QEP is strongly architecture-dependent. Rather than acting as a uniform accuracy amplifier, the QEP reshapes inference behavior in a manner determined by its interaction with the tensor-network frontend. In the present setting, TTN exhibits the most favorable overall profile, while MPS, TTN, and MERA occupy distinct operating regimes in the joint behavior of fusion dynamics and quantum-branch dispersion. The broader conclusion is therefore that learning performance is governed not by the quantum module alone, but by the co-design of tensor-network structure and post-aggregation quantum refinement.

The additional qubit-scaling and noise analyses further clarify the practical behavior of the QEP. For the present 64-dimensional latent input, performance becomes more stable from \(8\) qubits onward, while noisy conditions at \(8\) qubits degrade performance relative to the noiseless case in a noise-type-dependent manner. These results do not establish hardware-level benefit, but they help characterize the operating range and robustness of the hybrid quantum branch within the proposed pipeline.

From the systems perspective, the MPC benchmark shows that the dominant communication factor is the size of the protected latent representation. This highlights a second contribution of the architecture: tensor-network compression not only shapes the latent space on which quantum refinement operates, but also reduces the communication overhead to which secure computation is applied. Overall, the results support a co-design view of privacy-aware hybrid medical AI in which representation compression, post-aggregation quantum refinement, and secure deployment should be optimized jointly rather than treated as isolated components.

While the present results are simulator-based, an important long-term motivation of the proposed framework is future operation in regimes where classical simulation of the quantum branch is no longer practical. In such settings, post-aggregation quantum processing on tensor-network-compressed latent features may become more meaningful than direct quantum processing of the original high-dimensional input, because compression reduces the representation to a scale that is more compatible with limited quantum hardware while preserving task-relevant structure. From this perspective, the present study should be viewed as a systems-oriented step toward hybrid quantum federated learning under realistic privacy and resource constraints, rather than as a claim of near-term quantum advantage.

Future work will proceed in three directions. First, hardware validation in qubit regimes beyond practical classical simulation will be important for testing whether the proposed TN+QEP interface remains useful beyond the simulator-based setting considered here. Second, evaluation on real medical data will be needed to determine how far tensor-network frontends can compress clinically relevant inputs while preserving task-relevant structure and improving the practicality of downstream secure aggregation. Third, end-to-end implementation of the MPC pipeline, including explicit fixed-point execution, will be necessary to validate the present communication-cost model under realistic deployment conditions, assess the full system behavior of the proposed framework, and, when the post-aggregation quantum stage is externally delegated, examine privacy-preserving delegated quantum-computation mechanisms such as blind quantum computing and related protocols~\cite{Broadbent2009UBQC,Fitzsimons2017PrivateQC,Ma2022QEnclave}.

\section*{Acknowledgments}
The authors acknowledge helpful conversations with Mark Medum Bundgaad, Miyoji Kakinuki, Akiko Kamigori, Kazufumi Okazaki, Takuya Hasegawa, Toshiki Yasuda, Eriko Kaminishi, Tomah Sogabe, Naoki Yamamoto.
Hiroshi Yamauchi acknowledges the support of Yoshi-aki Shimada, Yosuke Komiyama and Ryuji Wakikawa.
This work was partly supported by NEDO Challenge Quantum Computing “Solve Social Issues!”. 
Portions of the manuscript were initially prepared in the authors’ native language and subsequently translated into English using AI-based translation tools. All content was subsequently reviewed and approved by all authors.
% ===================== References =====================
\bibliographystyle{IEEEtran}
\bibliography{bib}

\end{document}